\documentclass[%
twocolumn,
superscriptaddress,
 amsmath,amssymb,
 aps,prl
]{revtex4-2}

\usepackage{graphicx}
\usepackage{dcolumn}
\usepackage{bm}
\usepackage{color}
\usepackage{ulem}
\usepackage{xspace}
\usepackage{multirow}
\usepackage{hyperref}
\usepackage{physics}
\usepackage{braket}
\usepackage{here}
\usepackage[caption=false]{subfig}

\definecolor{green}{rgb}{0,0.6,0.1}

\begin{document}

\preprint{APS/123-QED}

\title{Origin of anomalous temperature dependence of Nernst effect in narrow-gap semiconductors}

\author{Ryota Masuki}
\email{masuki-ryota774@g.ecc.u-tokyo.ac.jp}
\author{Takuya Nomoto}
\affiliation{
Department of Applied Physics, The University of Tokyo,7-3-1 Hongo, Bunkyo-ku, Tokyo 113-8656
}

\author{Ryotaro Arita}
\affiliation{
Department of Applied Physics, The University of Tokyo,7-3-1 Hongo, Bunkyo-ku, Tokyo 113-8656
}
\affiliation{ 
RIKEN Center for Emergent Matter Science, 2-1 Hirosawa, Wako, Saitama 351-0198, Japan 
}


\date{\today}

\begin{abstract}
Based on the Boltzmann transport theory, we study the origin of the anomalous temperature dependence of the Nernst coefficient ($\nu$) due to the phonon-drag mechanism.
For narrow-gap semiconductors, we find that there are two characteristic temperatures at which a noticeable peak structure appears in $\nu$. Contrarily, the Seebeck coefficient ($S$) always has only one peak.
While the breakdown of the Sondheimer cancellation due to the momentum-dependence of the electron relaxation time is
essential for the peak in $\nu$ at low $T$, the contribution of the valence band to the phonon-drag current is essential for the peak at higher $T$. 
By considering this mechanism, we successfully reproduce $\nu$ and $S$ of FeSb$_2$ for which a gigantic phonon-drag effect is observed experimentally. 
\end{abstract}

\maketitle


{\it Introduction.}
The thermoelectric effect has been extensively studied for various materials due to its versatile potential applications such as power generation, energy conversion, and temperature sensing~\cite{shi2016recent, alam2013review, tomczak2018thermoelectricity}. While the longitudinal Seebeck effect has been exploited in many thermoelectric devices, those using the transverse Nernst effect are of great interest since they have many advantages: One can design flexible structures covering a heat source with a scalable generation of a large thermopower and high energy conversion efficiency~\cite{norwood1963theory,yang2017scalable,sakuraba2016potential,ikhlas2017large,Fe3Al,Co2MnGa}.

The thermoelectric effect is usually governed by the diffusion of electrons and becomes monotonically weak below the room temperature~\cite{he2013high}. However, in some situations, there can be another significant contribution at low temperature ($T$), which is called the phonon-drag effect. In the presence of strong electron-phonon interaction, a momentum transfer from nonequilibrium phonons to electrons occurs, and the thermopower can be dramatically enhanced~\cite{gurevich1945thermoelectric}.
The phonon-drag effect is particularly pronounced in some semiconductors with a long phonon lifetime, and 
is regarded as 
a promising mechanism to make high-performance thermoelectrics below the room temperature~\cite{zhou2015ab}.

The quantitative description of phonon drag is a fascinating problem to explore, having a long history~\cite{herring1954theory,frederikse1953thermoelectric, geballe1954seebeck, geballe1955seebeck, herring1958phonon, dunstan1974nernst, gurevich1989electron}. Regarding the Seebeck effect, intensive studies have been performed for several semiconductors in a wide range of $T$ and carrier concentrations~\cite{mahan2014seebeck, zhou2015ab, fiorentini2016thermoelectric, matsuura2019effect}.
On the other hand, it is not fully understood in which materials and in which conditions the phonon-drag contribution particularly enhances the Nernst coefficient. 
One representative example is FeSb$_2$, for which a huge phonon-drag effect is observed. Interestingly, it has been shown that its Nernst effect exhibits a characteristic $T$ dependence with multiple peaks, which are absent in the Seebeck coefficient~\cite{sun2009fesb, sun2013highly}. 
Recently, it has been proposed that this distinct difference in the longitudinal and transverse thermoelectric effect can be understood by the phonon-drag coupling to multiple in-gap states~\cite{battiato2015unified}.

In this Letter, 
we 
investigate another mechanism
of anomalous $T$ dependence of the Nernst effect
in narrow-gap semiconductors.
We first derive an electron-phonon coupled quantum Boltzmann equation
based on the Keldysh formalism and calculate the Seebeck ($S$) and Nernst coefficient ($\nu$)~\cite{keldysh47diagram, prange1964transport, luttinger1964theory}. 
In the approximation with a constant electron relaxation time (the so-called constant-$\tau$ approximation), $\nu$ vanishes at low $T$ due to the Sondheimer cancellation. However, if we take account of the momentum dependence of $\tau$, a noticeable peak structure appears. On top of that, if the bandgap $\varepsilon_g$ is sufficiently small,  another peak originating from the valence band appears at higher $T$. This result suggests that when $\tau$ has 
momentum dependence in a narrow-gap semiconductor, $\nu(T)$ will have a characteristic double-peak structure. On the other hand, we find that $S(T)$ always has one featureless peak, even if we go beyond the constant-$\tau$ approximation.

We then examine whether this mechanism plays a crucial role in the gigantic phonon-drag effect in FeSb$_2$.
FeSb$_2$ is a correlated narrow-gap semiconductor with a large effective mass~\cite{tomczak2018thermoelectricity, sun2013highly, tomczak2010thermopower} and exhibits a remarkably large Seebeck 
and Nernst effect
at cryogenic temperatures~\cite{sun2009huge, bentien2007colossal, sun2010narrow, sun2013highly, takahashi2016colossal}.
The maximum value of $|S|$ and $|\nu|$ reaches 45 mV/K at 10K~\cite{bentien2007colossal} and 3.2 mV/(KT) at 7K~\cite{sun2013highly}, respectively. While electron-correlation in FeSb$_2$ is considerably strong~\cite{jie2012electronic, tomczak2010thermopower, sun2013highly, sun2009fesb, petrovic2005kondo, takahashi2011low,PhysRevResearch.2.023190}, it has been recently shown that phonon drag is responsible for the colossal thermopower in this compound~\cite{tomczak2010thermopower,takahashi2016colossal, battiato2015unified, matsuura2019effect}.

For $S(T)$, we show that the present Boltzmann approach gives a result consistent with that obtained by an elaborate microscopic calculation based on linear response theory~\cite{ogata2019range, matsuura2019effect}. $S(T)$ has a single peak, regardless of whether or not the momentum dependence of $\tau$ is taken into account. On the other hand, we find that $\nu(T)$ has a  double peak structure when we go beyond the constant-$\tau$ approximation. The result shows a good agreement with the experiment~\cite{sun2013highly}. Our result indicates the importance of the momentum dependence of $\tau$ in narrow-gap semiconductors, which will be a useful guiding principle to control the phonon-drag effect and design efficient thermoelectric devices.

{\it Method.}
Starting from the Keldysh formalism~\cite{prange1964transport}, we derive the Boltzmann transport equation that includes the effect of impurity state up to linear order  
of impurity concentration.
The numerical calculation shows that the effect of impurity state on $S$ and $\nu$ is negligible 
within our approximation.
Hence, we start from the conventional Boltzmann transport equation that disregards the effect of the impurity state to simplify the discussion here.
Details of the treatment that considers the effect of impurity state is explained in the supplemental materials~\cite{supplement}. 

The phonon-drag effect on the thermoelectric effect is calculated from the Boltzmann transport equation in a method similar to that by Cantrell and Butcher~\cite{cantrell1987calculation}.
In the following,
we show a result for a simple case where one electron band 
and one phonon mode are involved.
We neglect the effect of the nonequilibrium electron distribution on the phonon distribution function. 
The change of the electron distribution function due to the phonon-drag effect is 
\begin{eqnarray*}
  [\delta f_{\text{qp}}(\bm{k})]_{\text{ph-drag}}^{\text{Seebeck}}
  &=& 
  -\tau_{\text{el,}\bm{k}} 
  \int \frac{d^3k'}{(2\pi)^3} \frac{d^3q}{(2\pi)^3} 
  \frac{\tau_{\text{ph}} \hbar \omega_{\bm{q}}}{k_B T^2} \\
  &&\times
  (\nabla_{\bm{q}} \omega_{\bm{q}} )\cdot (\nabla T)
  (P^{\bm{q}}_{\bm{k}\bm{k}'} - P^{\bm{q}}_{\bm{k}'\bm{k}} ),
\end{eqnarray*}
where 
\begin{eqnarray*}
  P^{\bm{q}}_{\bm{k}\bm{k}'} 
  &=&
  \frac{2\pi}{\hbar^2} |g_{\bm{q}}|^2 (1- f^{\text{eq}}_{\text{qp}}(\bm{k}) )f^{\text{eq}}_{\text{qp}}(\bm{k}') N^{\text{eq}}_{\text{ph}}(\bm{q})
  \\&&\times
  \delta(\omega_{\bm{q}} - \frac{1}{\hbar}(\varepsilon_{\bm{k}} - \varepsilon_{\bm{k}'}))
  (2\pi)^3 \delta(\bm{k}-\bm{k}'-\bm{q})
\end{eqnarray*}
is the transition amplitude of scattering from momentum $\bm{k}'$ to $\bm{k}$ mediated by phonon with momentum $\bm{q}$.
$f_{\text{qp}}$ and $N_{\text{ph}}$ are the distribution function of electrons at the quasiparticle peak and of the phonons respectively.
$\tau_{\text{el,}\bm{k}}$ is the relaxation time of electron excitation of momentum $\bm{k}$. 
$\tau_{\text{ph}}$ is the phonon lifetime.
$\varepsilon_{\bm{k}}$ and $\omega_{\bm{q}}$ represent the dispersions of the electron band and the phonon mode respectively.
$g_{\bm{q}}$ is the electron-phonon coupling constant.
Using this expression for $[\delta f_{\text{qp}}(\bm{k})]_{\text{ph-drag}}^{\text{Seebeck}} $, the phonon-drag contribution to the Seebeck coefficient can be calculated as
\begin{equation*}\begin{split}
  S_{\text{ph-drag}} = - \frac{g_s}{\sigma} \int \frac{d^3k}{(2\pi)^3} (-e{v}_{\bm{k}x}) \frac{[\delta f_{\text{qp}}(\bm{k})]_{\text{ph-drag}}^{\text{Seebeck}}}{(\nabla_x T)},
\end{split}\end{equation*}
where $\sigma$ is the electric conductivity 
and $g_s(=2)$ is the spin degeneracy.
Note that hereafter the temperature gradient is assumed to be in the $x$ direction.
In order to calculate the Nernst coefficient, we consider the case where external magnetic field is applied in the $z$ direction and retain the terms that are first order both in $\nabla T$ and in $\bm{B}$. 
The change of distribution function that contributes to the phonon-drag effect on the Nernst coefficient is 
\begin{equation*}\begin{split}
  [\delta f_{\text{qp}}(\bm{k})]_{\text{ph-drag}}^{\text{Nernst}}
  &= 
  \tau_{\text{el,}\bm{k}}\frac{1}{\hbar}e(\bm{v}_{\bm{k}} \times \bm{B}) \cdot \nabla_{\bm{k}}[\delta f_{\text{qp}}(\bm{k})]_{\text{ph-drag}}^{\text{Seebeck}}.
\end{split}\end{equation*}
The transverse component of the linear response coefficient $L_{12,yx}$ is calculated as
\begin{equation*}\begin{split}
  L_{12,yx} 
  &=
  \frac{g_s}{-(\nabla_x T)/T} \int \frac{d^3k}{(2\pi)^3} (-e v_{\bm{k}y}) [\delta f_{\text{qp}}(\bm{k})]_{\text{ph-drag}}^{\text{Nernst}},
\end{split}\end{equation*}
where the linear response tensor 
$L_{12}$ is defined by the relation $\bm{j} = \sigma \bm{E} + L_{12} (-\nabla T)/T$. 
Then, the Nernst coefficient is written as
\begin{equation}\begin{split}
  \nu 
  &= 
  S \frac{1}{B}\left( - \frac{\sigma_{yx}}{\sigma_{xx}} + \frac{L_{12,yx}}{L_{12,xx}} \right),
  \label{Nernst_general_formula}
\end{split}\end{equation}
using the linear response coefficients in the weak field limit. 
${L_{12,yx}}, L_{12,xx}$ can be calculated from the above discussion and $\sigma_{xx}, \sigma_{yx}$ can be calculated within the framework of the conventional Boltzmann transport equation~\cite{Ashcroft76}.

For typical semiconductors, we consider one conduction band, one valence band, and one longitudinal acoustic (LA) mode.
We assume that the conduction and the valence band have isotropic quadratic dispersions, and the LA phonon has an isotropic linear dispersion with phonon velocity $c_L$.
We use $|g_q|^2 = \frac{\hbar E_d^2}{2\rho c_L} q$ for the electron-phonon coupling constant, where $E_d$ is the deformation potential and $\rho$ is the mass density.
It is possible to analytically carry out all the angular integration under these assumptions and rewrite $L_{12,xx}$ and $L_{12,yx}$ by single integrals as
\begin{eqnarray}
  L^{\text{cond}}_{12,xx} &=& - \frac{e\hbar}{3\pi^2 m_{\text{c}}} \int_0^\infty dk k^3 \tau_{\text{el,c,}k} F_{\text{c,ph}}(k),\label{L12xx}\\
  L^{\text{cond}}_{12,yx} 
  &=&
  -B_z\frac{e^2 \hbar}{3\pi^2 m_{\text{c}}^2} \int_0^\infty dk k^3 \tau_{\text{el,c,}k}^2 F_{\text{c,ph}}(k),\label{L12yx}
\end{eqnarray}
for the conduction band contribution, where $m_{\text{c}}$ and $\tau_{\text{el,c,}k}$ are the effective mass and the electron relaxation time for the conduction band, respectively. The subscript c stands for the conduction band.
$F_{\text{c,ph}}(k)$ is defined as
\begin{equation}\begin{split}
  F_{\text{c,ph}}(k) = \frac{[\delta f_{\text{qp,c}}(\bm{k}=(k,0,0))]_{\text{ph-drag}}^{\text{Seebeck}}}{\tau_{\text{el,c},k}(-\nabla_x T)/T}.
  \label{F_cph}
\end{split}\end{equation}
where $\delta f_{\text{qp,c}}(\bm{k} = (k,0,0))$ is the change of electron distribution function in the conduction band evaluated at momentum $(k_x,k_y,k_z) = (k,0,0)$. The valence band contribution to $L_{12}$ tensor ($L_{12,xx}^{\rm val}$, $L_{12,yx}^{\rm val}$) can be written in a similar expression~\cite{supplement}.
The electric conductivity tensor can be calculated as
\begin{eqnarray}
  \sigma^{\text{cond}}_{xx} &=& - \frac{e \hbar}{3\pi^2  m_{\text{c}}}  \int_0^\infty dk k^3 \tau_{\text{el,c,}k} F_{\text{c,}E}(k),
  \label{sigma_cond_xx}\\
  \sigma^{\text{cond}}_{yx} &=& -B_z \frac{e^2 \hbar}{3\pi^2m_{\text{c}}^2}  \int_0^\infty dk k^3 \tau_{\text{el,c,}k}^2 F_{\text{c,}E}(k),
  \label{sigma_cond_yx}
\end{eqnarray}
where $F_{\text{c,}E}(k)$ is the electric field correspondent of $F_{\text{c,ph}}(k)$, which is defined as
\begin{equation}
\begin{split}
  F_{\text{c,}E}(k) 
  &= 
  \frac{[\delta f_{\text{qp,c}}((k,0,0))]_{E}}{E_x \tau_{\text{c,}\bm{k}}}
  =
  ev_{\text{c,}k} \left( \frac{\partial f^{\text{eq}}_{\text{qp,c}}}{\partial \varepsilon_{\text{c,}k}} \right).
  \label{F_cE}
\end{split}
\end{equation}
$[\delta f_{\text{qp,c}}]_{E}$ is the change of the distribution function induced by the external electric field. 
The valence band term of the electric conductivity tensor $\sigma^{\text{val}}$ can be calculated in an analogous way~\cite{supplement}.

The chemical potential is determined from the condition of charge neutrality, where the bandgap $\varepsilon_g$ and the donor binding energy $\varepsilon_b$ play a crucial role. 
Note that the electron concentration in the impurity state is $n_{\text{d}} \times \frac{1}{e^{\beta(\varepsilon_{\text{c,k=0}}-\varepsilon_b-\mu)}/2+1}$, which is apparently different from the Fermi-Dirac statistics because spin up and spin down electron cannot occupy the same impurity site at the same time due to the strong Coulomb repulsion~\cite{Ashcroft76}.

We consider the impurity scattering and the phonon scattering process to determine $1/\tau_{\text{el,c},\bm{k}}=1/\tau_{\text{c,imp},\bm{k}}+1/\tau_{\text{c,ph},\bm{k}}$.
For the impurity scattering, we adopt the Brooks-Herring model~\cite{chattopadhyay1981electron}, in which a Yukawa-type screened Coulomb potential is employed to represent the impurity potential. 
\begin{eqnarray*}
  \frac{1}{\tau_{\text{c,imp,}k}} 
  =
  \frac{4\pi \varepsilon_b n_{\text{d}} }{\hbar k^3} \int_0^2 \frac{xdx}{(x + q_D^2/(2k^2))^2}.
\end{eqnarray*}
where $n_{\text{d}}$ is the donor concentration and $q_D^2 = \frac{n_{\text{d}} e^2}{\epsilon k_B T}$. 
The dielectric constant $\epsilon$ is calculated from $\varepsilon_b$ and $m_c$
as $\epsilon = \epsilon_0 \times \sqrt{\frac{13.6\text{\ eV}}{\varepsilon_b} \frac{m_e}{m_{\text{c}}}}$. 
The electron scattering rate by the phonon can be calculated from the Boltzmann transport equation as
\begin{eqnarray*}
  \frac{1}{\tau_{\text{c,ph,}\bm{k}}}
  =
   \left( -k_B T \frac{\partial f^{\text{eq}}_{\text{qp,c}}}{\partial \varepsilon_{\bm{k}}} \right)^{-1}
  \int \frac{d^3k'}{(2\pi)^3} \frac{d^3q}{(2\pi)^3} 
  (P^{\bm{q}}_{\text{c},\bm{k}\bm{k}'} + P^{\bm{q}}_{\text{c},\bm{k}'\bm{k}}).
  \label{electron_phonon_electron_lifetime1}
\end{eqnarray*} 
Here, $P^{\bm{q}}_{\text{c},\bm{k}\bm{k}'}$ is the phonon mediated transition amplitude in the conduction band.
We assume that 
the hole relaxation time for the valence band is equal to that of the conduction band.

{\it Results and Discussion.}
Let us start with model calculations for the following two representative cases: A semiconductor with a moderate bandgap $\varepsilon_g=1.0$ eV and a narrow bandgap $\varepsilon_g=30$ meV.
For $\varepsilon_b$,
we use 50 meV for the former and 5 meV for the latter. The other parameters are the same for both cases. 
Namely, we set $n_{\text{d}} = 1.0\times10^{-8}$ \AA$^{-3}$, $c_L = 5000$ m/s, $E_d = 1.0$ eV, and $\rho = 5.0$ g/cm$^3$.
We 
assume that the valence and conduction band have the same effective mass (=$m_e$). 
We set $\tau_{\text{ph}}(T) = 1.0\times 10^{-7}\times 10^{- T/T_*}$ s, where $T_*= 20$ K. The parameters are in the same order as the fitting curve of the previous calculation result~\cite{matsuura2019effect}.

\begin{figure}[tb]
\vspace{0cm}
\begin{center}
\includegraphics[width=0.50\textwidth]{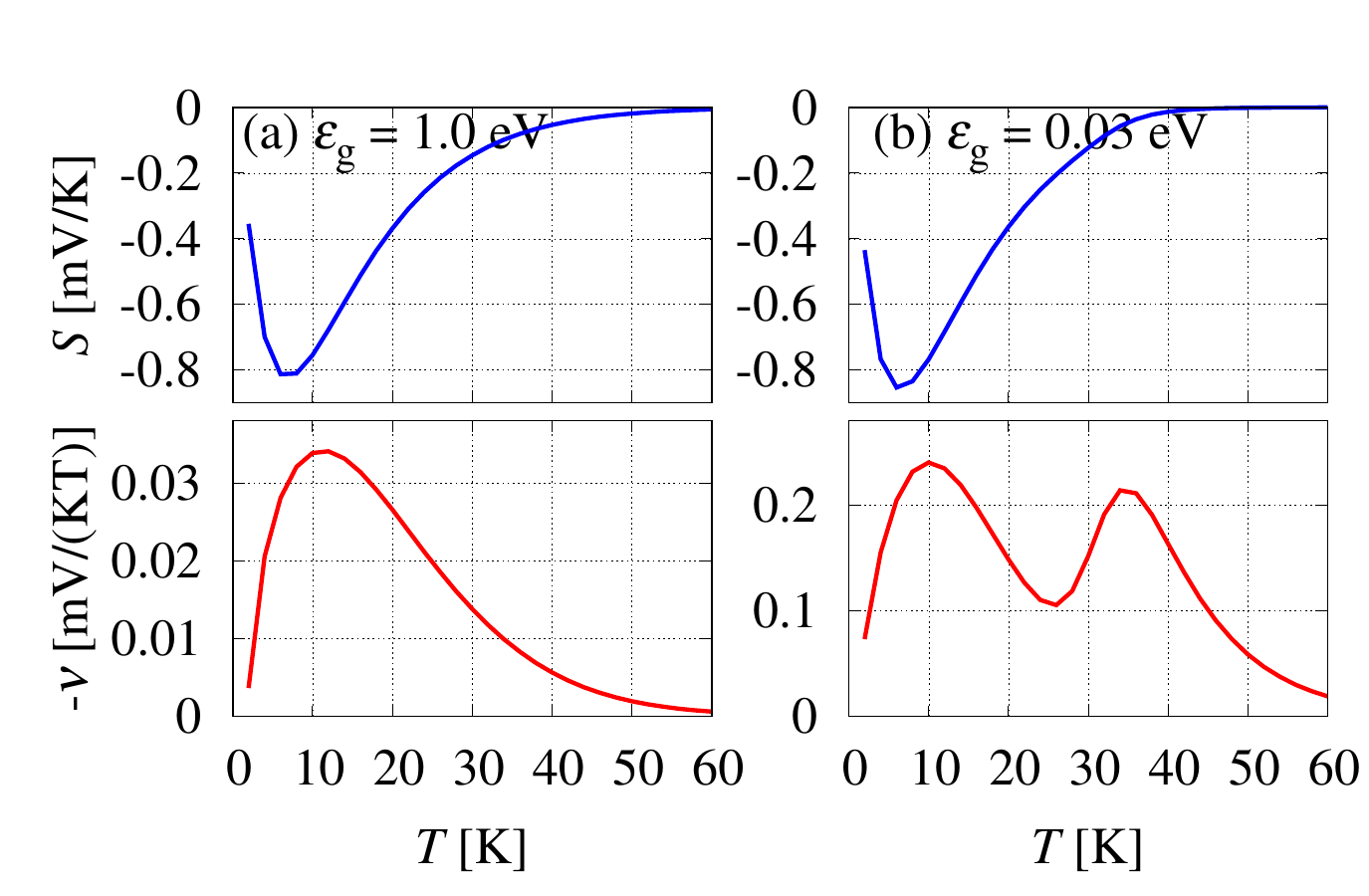}
\caption{
Phonon-drag contribution to the Seebeck coefficient $S$ and the Nernst coefficient $\nu$ for (a) a moderate-gap semiconductor and (b) a narrow-gap semiconductor. We set the bandgap $\varepsilon_g$ and the donor binding energy $\varepsilon_b$ as (a) $\varepsilon_g = 1.0$ eV and $\varepsilon_b = 0.05$ eV and (b) $\varepsilon_g = 30$ meV and $\varepsilon_b = 5$ meV. For other parameters, see the text. 
}
\label{model_SN}
\end{center}
\end{figure}

In Fig.~\ref{model_SN}, we show $S(T)$ and $\nu(T)$ for the moderate-gap case (a) and narrow-gap case (b).
For $S(T)$, we see that there is only one peak around $T=$ 10 K for both (a) and (b). On the other hand, while $\nu(T)$ has only one peak for (a) but two peaks for (b). Namely, on top of the peak around $T=$ 10 K, one additional peak appears around 40 K. 

The origin of these two peaks in $\nu(T)$ can be understood in terms of Eq.~(\ref{Nernst_general_formula}).
When $T$ is sufficiently lower than $\varepsilon_g$, we can neglect the contribution of the valence band and Eq.~(\ref{Nernst_general_formula}) is simplified as
\begin{eqnarray}
  \nu &=&
  \frac{S}{B_z} \left( - \frac{\sigma^{\text{cond}}_{yx} + \sigma^{\text{val}}_{yx}}{\sigma^{\text{cond}}_{xx} + \sigma^{\text{val}}_{xx}} + \frac{L^{\text{cond}}_{12,yx} + L^{\text{val}}_{12,yx}}{L^{\text{cond}}_{12,xx} + L^{\text{val}}_{12,xx}} \right) 
  \label{Nernst_general_formla_cond_val}
  \\
  &\simeq&
  \frac{S}{B_z} \left( - \frac{\sigma^{\text{cond}}_{yx}}{\sigma^{\text{cond}}_{xx} } + \frac{L^{\text{cond}}_{12,yx}}{L^{\text{cond}}_{12,xx} } \right).
  \label{Nernst_general_formla_cond}
\end{eqnarray}
It should be noted that we can show that
\begin{equation*}
\frac{\sigma^{\text{cond}}_{yx}/B_z }{\sigma^{\text{cond}}_{xx} }
=
\frac{L^{\text{cond}}_{12,yx}/B_z }{L^{\text{cond}}_{12,xx} }
=
\frac{e \tau_{\text{el,c}}}{m_{\text{c}}}
\end{equation*}
in the constant-$\tau$ approximation, so that the first and second term in Eq.~(\ref{Nernst_general_formla_cond}) cancel with each other (the so-called Sondheimer cancellation) and eventually $\nu$ becomes negligibly small.

\begin{figure}[tb]
\vspace{0cm}
\begin{center}
\includegraphics[width=0.5\textwidth]{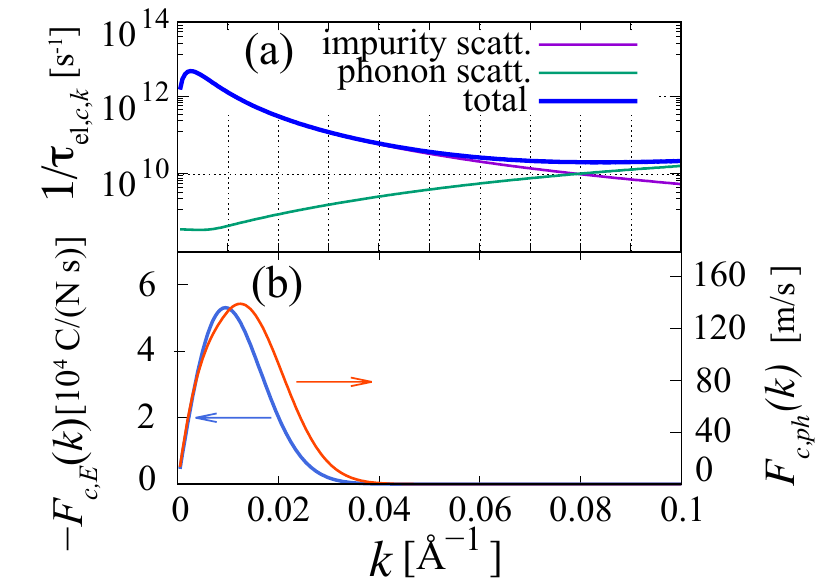}
\caption{
Momentum-dependence of (a) electron scattering rate and (b) 
$F_{\text{c,}E}(k)$ and $F_{\text{c,ph}}(k)$ (see Eqs.~(\ref{F_cph}) and (\ref{F_cE}) in the text)
for the narrow-gap semiconductor in Fig.~\ref{model_SN}(b) at 8 K.
}
\label{kdependence}
\end{center}
\end{figure}

However, if we consider the momentum dependence of $\tau_{{\rm el,c},{\bm k}}$,
this cancellation does not happen and $\nu(T)$ can be finite even at low $T$. 
In Fig.~\ref{kdependence}, we show $1/\tau_{{\rm el,c},{\bm k}}$, $F_{\text{c,}ph}(k)$ and $F_{\text{c,}E}(k)$ as a function of $k(=|{\bm k}|)$. 
We see that 
the peak of $F_{\text{c,}ph}(k)$ extends to $k$ larger than that of $F_{\text{c,}E}(k)$,
for which $\tau_{{\rm el,c},{\bm k}}$ is longer. In this situation, the second term in Eq.~(\ref{Nernst_general_formla_cond}) dominates over the first term (see Eqs.~(\ref{L12xx}),~(\ref{L12yx}),~(\ref{sigma_cond_xx}),~(\ref{sigma_cond_yx})).
Therefore, the momentum dependence of $\tau_{{\rm el,c},{\bm k}}$ is crucial for the formation of the peak in $\nu(T)$ around 10 K.

Let us now move on to the case of higher $T$.
As is discussed in the supplemental materials~\cite{supplement}, 
when $\varepsilon_g=5$~meV and $T\sim$ 30 K, 
the contribution of the valence band to the linear response coefficient (i.e., $\sigma^{\rm val}$ and $L^{\rm val}$) becomes comparable to that of the conduction band (i.e., $\sigma^{\rm cond}$ and $L^{\rm cond}$). 
As we can see from 
Table \ref{sign_table},
$\sigma_{yx}^{\rm cond}$
and 
$\sigma_{yx}^{\rm val}$
have opposite signs.
Contrarily, $L_{12,yx}^{\rm cond}$ 
and $L_{12,yx}^{\rm val}$ have the same sign, so that the second term in the r.h.s of Eq.~(\ref{Nernst_general_formla_cond_val}) becomes dominantly larger than the first term at $T\sim 30$ K, and eventually, $-\nu(T)$ becomes large. However, it should be noted that $\nu(T)$ vanishes in the limit of high $T$ since $S(T)$ becomes negligibly small. Thus $\nu(T)$ has a peak at intermediate $T\sim 30$ K. This mechanism does not work when $\varepsilon_g$ is larger than 1 eV $\sim 10^4$~K, so that the double peak structure in $\nu(T)$ appears only in narrow-gap semiconductors.
\begin{table}[htb]
  \caption{Signs of the conduction band and the valence band contribution to the linear response coefficients.}
  \begin{tabular}{|c|c|c|c|} \hline
    $\sigma_{xx}^{\text{cond}}$ & $\sigma_{xx}^{\text{val}}$ & $\sigma_{yx}^{\text{cond}}/B_z$ & $\sigma_{yx}^{\text{val}}/B_z$\\\hline
    + & + & + & $-$\\ \hline
    $L_{12,xx}^{\text{cond}}$ & $L_{12,xx}^{\text{val}}$ & $L_{12,yx}^{\text{cond}}/B_z$ & $L_{12,yx}^{\text{val}}/B_z$\\\hline
      $-$ & + & $-$ & $-$\\ \hline
  \end{tabular}
  \label{sign_table}
\end{table}

\begin{figure}[t]
\vspace{0cm}
\begin{center}
\includegraphics[width=0.48\textwidth]{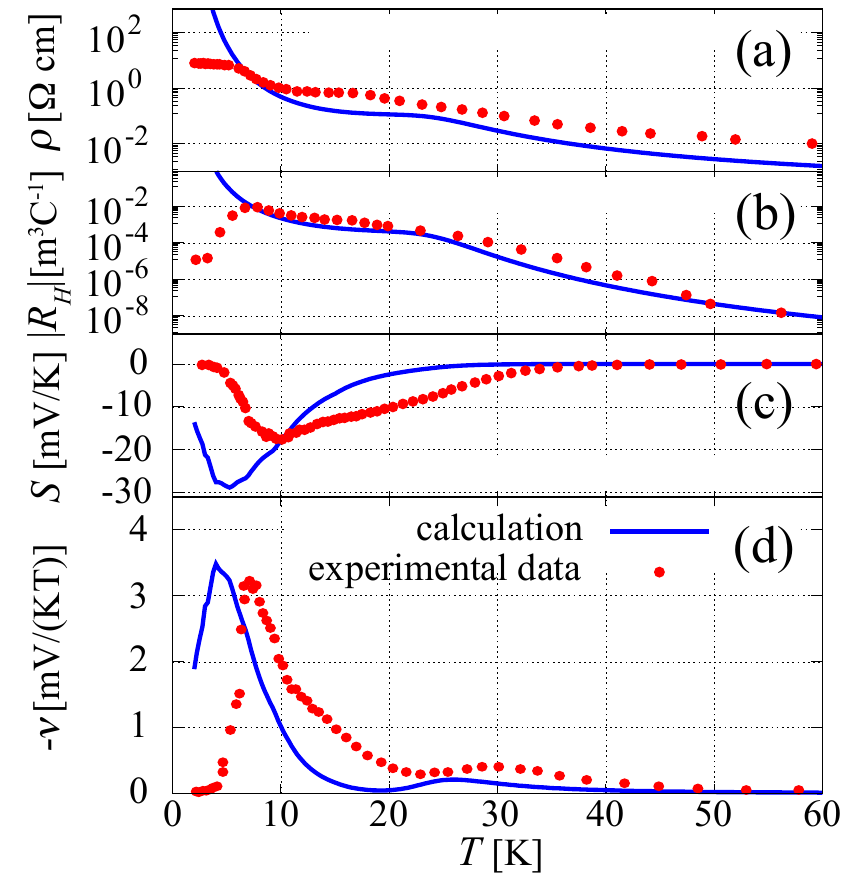}
\caption{
(a) Electric resistivity, (b) Hall coefficient, (c) Seebeck coefficient, and (d) Nernst coefficient of FeSb$_2$ compared with the experimental result~\cite{sun2013highly}.
}
\label{FeSb2_Sun2013_S_N}
\end{center}
\end{figure}

Finally, let us consider the case of FeSb$_2$. 
We show the result for $S(T)$ and $\nu(T)$ in Fig.~\ref{FeSb2_Sun2013_S_N}. 
In the calculation, the effective mass of the conduction and the valence band are set to $m_{\text{c}} = m_{\text{v}} = 5m_e$~\cite{matsuura2019effect,takahashi2016colossal}. The bandgap, impurity binding energy are set to $\varepsilon_g = $ 28 meV, $\varepsilon_b = $ 6 meV, respectively~\cite{sun2013highly}. 
We set $c_L = $ 3100 m/s, $ n_{\text{d}} = 1.9\times 10^{-8}$ \AA$^{-3}$.
The donor concentration $n_{\text{d}}$ is determined so that the carrier concentration at low $T$ be the same order as the experiment~\cite{sun2013highly}. The deformation potential is set to $E_d=0.85$ eV, which is a typical value for semiconductors. The mass density is set to $\rho$ = 8.2 g/cm$^3$.
The phonon lifetime $\tau_{\text{ph}}(T)$ is determined to reproduce the experimental result of the thermal conductivity~\cite{matsuura2019effect}.

We see that the result is in good agreement with the experiment~\cite{sun2013highly}.
Note that the result for $S(T)$ is also consistent with the previous result based on microscopic linear response theory~\cite{matsuura2019effect}.
The electric resistivity and the Hall coefficient reach a plateau around $T = 10$ K to 20 K. In this temperature range, a large fraction of the electrons in the impurity state is excited to the conduction band while the valence band is still almost completely occupied. 
Above this 
plateau, the carrier concentration increases due to excitation from the valence to the conduction band, which is consistent with the scenario for the anomalous $T$ dependence of $\nu(T)$ in narrow-gap semiconductors.
These results clarify that the origin of the huge phonon-drag contribution to $\nu$ of FeSb$_2$ at low temperatures is the momentum dependence of $\tau_{\text{el,}\bm{k}}$ and the small $\varepsilon_g$.

{\it Conclusion.}
We investigate the phonon-drag contribution to the Seebeck and the Nernst coefficient using the Boltzmann transport equation. 
We identify that the momentum-dependent electron relaxation time is crucial to the large negative Nernst coefficient at low temperature, which breaks the Sondheimer cancellation. 
Furthermore, we find that the Nernst coefficient has another peak at higher temperature if the bandgap is sufficiently small.
Considering this mechanism, we calculate the electric resistivity, the Hall coefficient, the Seebeck and the Nernst coefficient of FeSb$_2$, and succeed in reproducing the experimental result. 
Our results propose another possible origin of the anomalous temperature dependence of the Nernst coefficient of FeSb$_2$, which provides a possibility to utilize the phonon-drag effect to design good thermoelectric materials.

{\it Acknowledgements.}
We would like to thank J. M. Tomczak for his fruitful comments.
This work was supported by a Grant-in-Aid for Scientific  Research (No. 19K14654,  No. 19H05825, and No. 16H06345) from Ministry of Education, Culture, Sports, Science and Technology, and CREST (JPMJCR18T3) from the Japan Science and Technology Agency.

\bibliography{apssamp}

\end{document}